\documentclass[11pt]{article}

\input{macros}
\usepackage{nicefrac}
\usepackage{wrapfig}
\usepackage{microtype}

\newcommand{\Time}{\mathbb{T}}
\newcommand{\Utime}{\mathbb{U}}
\newcommand{\myAlgorithm}[1]{{\textup{\texttt{A}}}_\textit{{#1}}}
\newcommand{\Aold}{\myAlgorithm{universal}}
\newcommand{\Amem}{\myAlgorithm{memory}}
\newcommand{\Apareto}{\myAlgorithm{pareto}}
\newcommand{\Anew}{{\myAlgorithm{}^\star}}
\newcommand{\Acord}{\myAlgorithm{cord}}
\newcommand{\column}{C}
\newcommand{\Valid}{\mathcal{V}}
\newcommand{\Active}{\textup{\texttt{ac}}}
\newcommand{\dm}{\diff\mu}
\newcommand{\zoom}[3]{{#1}_{\overrightarrow{#2,#3}}}
\newcommand{\OPT}{\text{\tt OPT}}
\newcommand{\Cont}{\mathcal{F}}

\title{Parallel Search with no Coordination\thanks{This work has received funding from the European Research Council (ERC) under the European Union's Horizon 2020 research and innovation programme (grant agreement No 648032).}}

\author{Amos Korman}
\affil{CNRS and University Paris Diderot, Paris, France\\
\Email{amos.korman@irif.fr}}
\author{Yoav Rodeh}
\affil{Weizmann Institute, Rehovot, Israel\\
\Email{yoav.rodeh@gmail.com}}

\begin{document}

\maketitle

\begin{abstract}
We consider a parallel version of a classical Bayesian search problem. $k$ agents are looking for a treasure that is placed in one of the boxes indexed by $\Natural^+$ according to a known distribution $p$. The aim is to minimize the expected time until the first agent finds it. Searchers run in parallel where at each time step each searcher can ``peek'' into a box.
A basic family of algorithms which are inherently robust is  \emph{non-coordinating} algorithms. Such algorithms act independently at each searcher, differing only by their probabilistic choices. We are interested in the price incurred by employing such algorithms when compared with the case of full coordination. 

 We first show that there exists a non-coordination algorithm, that knowing only the relative likelihood of boxes according to $p$, has expected running time of at most $10+4(1+\frac{1}{k})^2 T$, where $T$ is the expected running time of the best fully coordinated algorithm. This result is obtained by applying a refined version of the main algorithm suggested by Fraigniaud, Korman and Rodeh in STOC'16, which was designed for the context of linear parallel search.

We then describe an optimal non-coordinating algorithm for the case where the distribution $p$ is known. The running time of this algorithm is difficult to analyse in general, but we calculate it for several examples. In the case where $p$ is uniform over a finite set of boxes, then the algorithm just checks boxes uniformly at random among all non-checked boxes and is essentially $2$ times worse than the coordinating algorithm.
We also show simple algorithms for Pareto distributions over $M$ boxes.  That is, in the case where $p(x) \sim 1/x^b$ for $0< b < 1$, we suggest the following algorithm: at step $t$ choose uniformly from the boxes unchecked in $\set{1, \ldots, \min(M, \Floor{t/\sigma})}$, where $\sigma = b/(b + k - 1)$. 
It turns out this algorithm is asymptotically optimal, and runs about $2+b$ times worse than the case of full coordination.  
\end{abstract}

\thispagestyle{empty}
\clearpage
\setcounter{page}{1}

\section{Introduction}

We consider a parallel variant of the classical Bayesian search problem, typically attributed to Blackwell \cite{blackwell}.  A treasure is placed in one of the boxes indexed by $\Natural^+$ according to some known distribution $p$. As $p$ is known, we can assume that the boxes are ordered so that $p$ is non-increasing. 
Denote $M = \max\stset{x}{p(x)>0}$, which can be $\infty$.
There are $k$ agents that search for the treasure, aiming to minimize the expected time until the first one finds it, where looking into a box takes one unit of time.
We shall assume that algorithms know the number of searchers $k$. 

If coordination is allowed, a simple application of the rearrangement inequality shows that letting agent $i$ peek into box $(t-1)k + i$ at time $t$ is an optimal algorithm (see Appendix \ref{apx:coordinating}). Denote this algorithm $\Acord$, and note that its expected running time is
$\sum_x p(x) \Ceil{x/k}$, giving a speedup of essentially $k$ compared to just one searcher. 
However,  as simple as this algorithm is, it is very sensitive to faults of all sorts.
For example, if one searcher crashes at some point during the execution then the searchers may completely miss the treasure, unless the protocol employs some mechanism for detecting such faults\footnote{It is actually an interesting and non trivial question to find efficient and robust algorithms that are allowed to coordinate \cite{FaultyRobots}. Of course, our non-coordinating algorithms fall under this category, but one may potentially improve the running time by allowing coordination.}.  

A class of search algorithms which is of particular interest  is \emph{non-coordinating} algorithms \cite{alon_many_2008,fraigniaud_parallel_2016}.  In such an algorithm, all searchers operate independently, executing the same protocol, differing only in the outcome of the flips of their random coins. 
With such a strong restriction on the coordination, one cannot expect that many search problems could be efficiently parallelized. However, when such a parallelization can be achieved,  the benefit can potentially be high, not only in terms of saving in communication and overhead in computation, but also in terms of robustness.  
To get some intuition, assume that an oblivious adversary is allowed to crash at most $f$ out of the $k$ searchers at arbitrary points in time during the execution. To overcome the presence of at most $f$ faults, one can simply run the non-coordinating algorithm that is designed for the case of $k-f$ searchers. If the running time of a non-coordinating algorithm without crashes is $T(k)$, then the running time of the new robust algorithm would be at most $T(k-f)$. 
This is because  the correct operation as well as the running time of a non-coordinating algorithm can only improve  if more searchers than planned are actually being used. Note that even when coordination is allowed, one cannot expect to obtain robustness at a cheaper price since the number of searchers that remain alive is in the worst case $k-f$.

For an algorithm $A$, and $k \geq 2$, 
denote by $\Time_k(A, x)$ the expected running time if the treasure is placed at $x$, when running algorithm $A$ with $k$ searchers. Note that by ``running time" we actually mean the expected number of boxes peeked into by each searcher, as we are mostly interested in query complexity.
Further, for a distribution $p$ over the boxes,
denote the expected time to find the treasure when it is placed in one of the boxes according
to $p$:
\[
\Time_{p, k}(A) = \sum_x p(x) \Time_k(A, x)
\]
In this notation, the expected running time of the optimal coordinating algorithm is $\Time_{p,k}(\Acord)$. We are interested in the 
connection between these two terms, and specifically in identifying non-coordinating algorithms that minimize the additive and multiplicative factors $a$ and $b$ such that:
\[
{\Time_{p, k}(A)}\leq a + b{\Time_{p,k}(\Acord)}
\]
We remark, for readability's sake, the subscripts above, as well as most subscripts in the text that follows, will be dropped when clear from context. Also, the number of agents $k \geq 2$ will be fixed and so omitted from formal statements. This will many times go for $p$ as well.
Also note that there are distributions where no algorithm can achieve finite running time, such as $p(x) = c/x^2$, where the expected placement of the treasure is unbounded. 
We shall therefore always assume that $\sum_x p(x)x < \infty$, and so, for example, $\Time(\Acord)$ is always defined. 

\subsection{Our Results}
We first show that there exists a simple and highly efficient  algorithm, denoted $\Aold$, which for large $k$ enjoys a multiplicative factor that tends to $4$.
In this algorithm, each agent, at phase $t$, checks two different uniformly chosen boxes of those it did not check yet in $\set{1, \ldots, t(k+1)}$.
This algorithm is universal in the sense that it does not depend on the details of the distribution $p$, and assumes only the knowledge of the relative likelihood of the boxes, that is, their order.
\rTheorem{thmUniversal}{
$
\Time(\Aold) \leq 10 + 4\B{1 - \R{k+1}}^2 \Time(\Acord)
$
}
Note that this gives improvement over the trivial one searcher for every $k$. Even for $k=2$ we get that for large enough $x$, this runs at $8/9$'s the time of the lone searcher.

Algorithm $\Aold$ remembers all the boxes it checked and so needs  memory which is linear in its running time. We also consider $\Amem$ which at phase $t$ chooses uniformly two boxes in $\set{1, \ldots, tk}$. This algorithm uses only logarithmic memory in its running time, and for large number of searchers performs almost as well:
\rTheorem{thmMemory}{
$
\Time(\Amem) \leq 2 + 4\Time(\Acord)
$
} 
Both algorithms $\Aold$ and $\Amem$ where actually given in \cite{fraigniaud_parallel_2016} to tackle the setting  of linear search  with an adversarially placed treasure. 
We note, however, that when applied in our context, the bounds established in \cite{fraigniaud_parallel_2016} only guarantee that the additive term is some unknown, possibly large constant. 
To prove that this constant is in fact small we had to refine the upper bound analysis of \cite{fraigniaud_parallel_2016}, and prove tighter bounds on the Gamma function. 

We next present Algorithm $\Anew$, that given access to the the exact distribution $p$ (and not only the order of the boxes), gives the optimal expected running time:
\rTheorem{thmOptimal}{
For every non-coordinating algorithm $A$,
$
\Time(\Anew) \leq \Time(A).
$
}
An interesting property satisfied by this algorithm, is that at any time during the execution,  all boxes that previously received a positive probability to be  checked, are now going to be checked with equal probability.

Calculating the running time of $\Anew$ can become challenging for specific distributions, and the rest of the paper shows a few interesting examples.
A simple one is when $p$ is the uniform distribution over a finite domain. In this case, running $\Anew$, at each step each agent chooses a box uniformly among those it did not check yet. This natural choice for an algorithm therefore turns out to be optimal, and yields a multiplicative factor of essentially $2$ when compared to $\Acord$. 

On the other extremity there are exponential distributions. Such distributions strongly concentrate the probability on the first few boxes, and so a good algorithm would invest in optimizing the parallel performance on a constant number of boxes. As we are concerned with non-trivial behavior over many boxes, we turn our attention to investigate Pareto distribution, which spread the distribution more gradually.

Specifically, we consider the family of Pareto distributions over $M$ boxes, thinking of $M$ as large. Here, for some $0 < b < 1$, 
for all $x\leq M$, $p(x) = I/x^b$, where $I$ is the normalization factor, and $p(x) = 0$ for larger $x$.
While $\Anew$ is optimal, it is quite complex and difficult to analyse. We present a simple algorithm $\Apareto$ that is asymptotically optimal. In $\Apareto$, at step $t$, an agent chooses uniformly from one of the boxes it did not check yet in $\set{1, \ldots, \min(M, \Floor{t/\sigma})}$, where $\sigma = b/(b + k - 1)$. 
\rTheorem{thmPareto}{
For $0 < b < 1$,    
$
\lim_{M \rightarrow \infty} 
\F{\Time_{r_{M,b}}(\Apareto)}{\Time_{r_{M,b}}(\Acord)} 
=
k\sigma(2-\sigma) + \F{k}{k+1}(2-b)(1-\sigma)^2
$.
Furthermore, no non-coordinating algorithm can achieve a better limit bound.
}
When $b$ is close to $1$, then $\sigma \approx 1/k$ and the factor becomes $(3k-1)/(k+1)$. 
For $k=2$ this is $5/3$ compared to $16/9$ achieved by $\Aold$, and for large $k$ this tends to $3$ as opposed to $4$.
For smaller $b$'s the result is not as clean, but assuming $k$ is large, then $\sigma \approx b/k$, and we get that the ratio is about $2 + b$.
This makes sense, as when $b$ approaches $0$, the distribution becomes uniform, where we already know that this factor is $2$ for large $k$. 

Finally, we note that most of our algorithms are very simple and hence applicable. From the technical point of view, our results  illustrate deep connections between the general probabilistic parallel search setting considered here, and the setting of parallel {\em linear} search studied in \cite{fraigniaud_parallel_2016}.

\subsection{Related work}

The study of parallel search by non-coordinating algorithms has recently been advocated by Fraigniaud, Korman and Rodeh as a simple way to obtain robustness while avoiding communication overheads \cite{fraigniaud_parallel_2016}. 
The setting therein, however, differs from ours by two fundamental characteristics: First, they assumed that the treasure is placed by an adversary. The second major difference is that they focused on a {\em linear search} setting (see also \cite{baezayates_searching_1993,Beck,Linear}), in which the boxes are linearly ordered and the objective is to find a treasure placed in a box in time that is compared to its index. That is, if the treasure is placed in index $x$, then the running time of the parallel algorithm should be compared to $x/k$. 
Although this linear search setting may seem somewhat specific compared to the setting studied in the current paper, it turns out that there are  important connections between the two settings, both in terms of techniques and results. See Section \ref{sec:reduction} for more details.

The case of a single searcher that searches for a randomly pleased treasure has receives significant amount of attention from the communities of statistics, operational research and computer science, see e.g., \cite{blackwell,Zamir,Chew}, and has been studied under various settings, 
including the case that there are different costs associated with queries, that queries can be noisy, and that the target may be mobile, see the book  \cite{book}. 
As we initiate its parallel version, we consider only the most basic form of the problem, yet, we note that most of our results can be extended to the case in which weighted costs are associated with queries.  

In general, when it comes to parallel search, most of the literature deals with mobile agents that search graphs of different topologies, and typically employ some form of communication between themselves. The literature on this subject is vast, and some good references can be found, e.g., in \cite{Games,Gal,Das,Flochini,Autonomous}. The major difference between our setting and the mobile agent setting, is that we allow ``random access'' to the different boxes. That is, our searcher can jump between different boxes at no cost.
In other words, our focus is on the {\em query complexity} rather than the {\em move complexity}.

Multiple random walkers are a special case of non-coordinating searchers. In a series of papers \cite{alon_many_2008, cooper_multiple_2009, elsasser_tight_2011,efremenko_how_2009} several  results regarding hitting time, cover time and mixing times are established, such as a linear speedup for several graph families including expanders and random graphs.
Non-coordinating searchers have also been studied in the context of the ANTS problem, a parallel variant of the cow-path problem on the grid \cite{baezayates_searching_1993,kao_searching_1993}, which was introduced in \cite{feinerman_memory_2012,feinerman_collaborative_2012} 
motivated by applications to central search foraging by desert ants. For example, it was shown in  \cite{feinerman_memory_2012,feinerman_collaborative_2012}  that a speedup of $O(k)$ can be achieved with $k$ non-coordinating searchers, and that a linear speedup cannot be achieved unless the agents have some knowledge of $k$. 

Finally, BOINC \cite{BOINC}  (Berkeley Open Infrastructure for Network Computing) is a platform for volunteer computing supporting dozens of projects including the famous SETI@home analyzing radio signals for identifying signs of extra terrestrial intelligence. Most projects maintained at BOINC use parallel  search mechanisms where a central server controls and distributes the work to volunteers. The framework in this paper is a potential abstraction for projects operated at platforms similar to BOINC with hundreds of thousands distributed searchers.

\section{Ordering of Boxes is Known}\label{sec:reduction}

In \cite{fraigniaud_parallel_2016}, the authors consider a somewhat different scenario. The boxes are ordered linearly by some predefined importance, and the treasure is placed in one of them by an adversary. In such a situation, a lone searcher will check the boxes according to their order, and so box $x$ will be checked by time $x$. 
They present algorithm $\Aold$, in which each agent, at phase $t$, checks two different uniformly chosen boxes of those it did not check yet in $\set{1, \ldots, t(k+1)}$.
It is shown there that:
\[
\limsup_{x \rightarrow \infty} \F{\Time(\Aold,x)}{x} = \F{4k}{(k+1)^2}
\]
and that it is in fact optimal in this way.
That is, in that setting, it has the best speedup compared to the lone searcher when taking large enough $x$.

If $\Aold$ would give this result for all $x$ and not only large ones, it will solve the case of a randomly placed treasure with surprising efficiency. All one has to do is set the importance of the boxes according to the likelihood of the treasure being placed there. 
The following claim is proved in Appendix \ref{apx:refinedUniversal} via a refined analysis of that done in \cite{fraigniaud_parallel_2016}, and shows that the limsup only hides a small additive term:
\rClaim{clmRefined}{
For all $x$,
$
\Time(\Aold,x) \leq 10 + \F{4k}{(k+1)^2} x
$.
}
A major ingredient in the proof is
the following lemma:
\rLemma{lmGamma}{For integers $b \geq a \geq 1$, and $0 < \phi \leq 1$,
$
\prod_{i=a}^b \F{i}{i+\phi} \leq \BF{a}{b}^\phi
$.
}
Using properties of the Gamma function it is easy to see that the
two sides of the equation are asymptotically equal, but this is not
enough to prove our result as we need the inequality for small $a$
and $b$ as well. Using Claim \ref{clmRefined} the following is straightforward:
\thmUniversal
\BPF
\[\BA
&
\Time(\Aold)
 = 
\sum_x p(x) \Time_k(\Aold, x)
\leq
10 + \F{4k}{(k+1)^2} \sum_x p(x) x
\\ & \leq 
10 + \F{4k^2}{(k+1)^2} \sum_x p(x)\Ceil{\F{x}{k}}
= 10 + 4\B{1 - \R{k+1}}^2 \Time(\Acord)
\EA\]
\EPF
At \cite{fraigniaud_parallel_2016}, the authors introduce a memory efficient version of $\Aold$, which we present here, slightly altered, as $\Amem$. In it,
each agent, at phase $t$, checks two uniformly chosen boxes of those in $\set{1, \ldots, kt}$.
The following is proved in Appendix \ref{apx:refinedMemory}:
\rClaim{clmRefinedMemory}{
For all $x$,
$
\Time(\Amem,x) \leq 2 + 4 \Ceil{\F x k}
$.
}
Note that for $k \leq 4$ this is of no use, as running the trivial one searcher will do better. This claim immediately proves,
\thmMemory
While $\Amem$ is not optimal as $\Aold$ is, as $k$ grows the difference between them grows smaller, and $\Amem$'s simplicity and efficiency make it an outstanding candidate for real life purposes.

\section{Exact Distribution is Known}

Ignoring the small additive term in Theorem \ref{thmUniversal}, as $k$ grows larger we get that $\Aold$ is about $4$ times worse than the best coordinating algorithm. In the remainder of the paper we show it is possible to improve on this if the exact distribution is known.

\subsection{Preliminaries}

Consider a non-coordinated algorithm $A$ that is running on $k$ agents.
Focusing on just one agent, denote by
$A(x,t)$ the probability that by time $t$, box $x$ was not already checked by this agent.
Hence, the probability that none of the $k$ agents checked $x$ by time $t$
is $A(x,t)^k$. In fact, as we shall soon see, the information encoded in this functional view of $A$ is all that is needed to assess its running time. 
First note:
\begin{observation} \label{obs:alg2mat}
The function corresponding to algorithm $A$ satisfies $A(x,0) = 1$ for all $x$. Also, for all $x$ and $t\geq 1$: 
\[
A(x, t) = 
A(x, t-1) \cdot
\cprob{\textup{\begin{tabular}{l}$x$ wasn't checked\\ at time $t$\end{tabular}}}
{\textup{\begin{tabular}{l}$x$ wasn't checked \\ prior to time $t$\end{tabular}}}
\]
\end{observation}

Let us now consider such functions on their own, possibly without a corresponding algorithm.
Let\footnote{The letter $N$ stands for ``probability of \textit{not} being checked up to time".} $N : \Natural^+ \times \Natural \rightarrow [0,1]$. 
For time $t$, denote:
\[
C_N(t) = \sum_x 1 - N(x,t)
\]
In the case of an algorithm $A$, $C_A(t)$ is the expected number of elements that were checked by time $t$ by just one of the searchers running $A$, and is therefore at most $t$. We say that $N$ satisfies the {\em column requirement} at time $t$ if $C_N(t) \leq t$. Also, define the set of {\em valid} functions as:
\[
\Valid = \stset{N : \Natural^+ \times \Natural \rightarrow [0,1]}{\forall t, \column_N(t) \leq t}
\]
and so functions corresponding to algorithms are always valid. Finally, the ``running time" of $N$:
\[
\Time_{p,k}(N) =
\sum_x p(x) \sum_t N(x,t)^k = \sum_t \sum_x p(x) N(x,t)^k
\]
The sum on $t$ is from $0$ to $\infty$, and these limits will be omitted whenever clear from context. 
This is clearly defined so that $\Time(A)$ is indeed the expected running time of algorithm $A$, 
as $\Time(A, x) = \sum_t \prob{x \text{ wasn't found by time } t} = \sum_t A(x,t)^k$.

To lower bound the running time of algorithms, we find the optimal $N \in \Valid$, in the sense that it minimizes $\Time(N)$. For that, we introduce a generalized version of the main Lemma of \cite{fraigniaud_parallel_2016} which we prove in Appendix \ref{apx:mainLemma}. At this point we only need a very simple version of the lemma, yet we present it in its full glory, as we will need it later in the paper. 
The current version improves on the original lemma of \cite{fraigniaud_parallel_2016} as it applies  to general measurable functions, instead of only continuous and bounded ones. 
In addition, the measure theoretic proof is much more elegant and concise than the original one. 
\subsection{Main Lemma}
The notation that follows is in measure theory style.
Fix some $k \geq 2$ and let $(X, \mathcal{X}, \mu)$ be a measure space. 
For $T \geq 0$, denote by $V(T)$ the set of measurable functions $f : X \rightarrow [0,1]$ such that $\int 1 - f \dm \leq T$.
For a measurable function $c: X \rightarrow [0,\infty)$, and $\alpha \geq 0$ define the function $f_{c, \alpha} : X \rightarrow [0,1]$ as:
\[\BA
f_{c, \alpha}(x) = \begin{cases}
1 & c(x) = 0 \\
\min\B{1, \alpha c(x)^{-\R{k-1}}} & \text{otherwise}
\end{cases}
\EA\]
\rLemma{lmMain}{
For a given $c$ and $T$ as above,
if there is some $h \in V(T)$ such that $\int c h^k \dm< \infty$,
then there exists $\alpha \geq 0$, such that $f_{c, \alpha} \in V(T)$, and for every $g \in V(T)$,
$\int c f^k_{c, \alpha} \dm \leq \int c g^k \dm$. Furthermore, this $\alpha$ is minimal among those satisfying $f_{c, \alpha} \in V(T)$.
}
Towards finding the optimal $N \in \Valid$, fix some $t$, and then $N \in \Valid$, means 
$\sum_x 1 - N(x,t) \leq t$, and the aim is to minimize $\sum_x p(x) N(x,t)^k$. As this can be done for each $t$ completely separately, Lemma \ref{lmMain} comes into play.

\begin{claim}
The following function $L$ is in $\Valid$, and achieves minimal $\Time(\cdot)$ over all valid functions.
\[
L_{p,k}(x,t) = \begin{cases}
1 & p(x) = 0 \\
\min(1, \alpha(t) q(x)) & \textup{\text{otherwise}}
\end{cases}
\]
Where $q(x) = p(x)^{-\R{k-1}}$, 
and for all $t$, $\alpha(t) \geq 0$ is the minimal such that $L_{p,k} \in \Valid$.
\end{claim}
\BPF
Fix $t$. Setting $X = \Natural^+$ with the trivial measure $\mu(x)=1$ for all $x$, $T=t$ and $c=p$, Lemma \ref{lmMain} gives the values of the optimal $N$ for this specific $t$. 
To check the condition of the lemma, take the constant function $h(x) = 1$. Clearly $h \in V(t)$, and 
$\int c h^k \dm = \sum_x p(x) = 1 < \infty$. 
\EPF

The following basically says that $L$, if thought of as an algorithm, never rechecks a box. See the proof in Appendix \ref{apx:obsExact}.
\rObservation{obsExact}{
For every $t<M$, $\column_L(t) = t$, and for $t \geq M$, $L(x,t) = 0$ everywhere.
}

\begin{wrapfigure}{r}{0.26\textwidth}
	\vspace{-20pt}
	\[
	\begin{array}{c|cccc}
	_{x\downarrow}^{t\rightarrow} & 0 & 1 & 2 & 3 \\
	\hline
	1 & 1  & 0.4  & \nicefrac{2}{11} & 0   \\
	2 & 1  & 0.6  & \nicefrac{3}{11} & 0  \\
	3 & 1  & 1   & \nicefrac{6}{11} & 0  \\
	\end{array}
	\]
\vspace{-20pt}
\end{wrapfigure}

As an illustration consider a simple example: $k=2$, $p(1) = 1/2$, $p(2) = 1/3$, and $p(3) = 1/6$. In this case, $q(1) = 2$, $q(2) = 3$ and $q(3) = 6$, and some quick calculations show that $\alpha(1) = 1/5$, $\alpha(2) = 1/11$, and $\alpha(3) = 0$.  
From these we get the matrix $L$ on the right. Note that Observation \ref{obsExact} holds, as the sum of column $t$ is indeed equal to $M-t$.

\subsection{Optimal Algorithm}

Although it may seem that every valid function $N$ has a corresponding algorithm, it is not at all clear, because the conditional probabilities arising from Observation \ref{obs:alg2mat} quickly become complicated for general $N$. However, it turns out that because of the specific structure $L$ has, there is in fact an algorithm that has it as its function.

For instance, a corresponding algorithm for the example above is:
(1) choose box $1$ w.p.\ $0.6$, and otherwise choose box $2$.
(2) choose box $3$ w.p.\  $5/11$, and otherwise the unchosen box of 1 and 2.
(3) choose the last remaining box.
Note especially step (2), where the remaining probability of $6/11$ is used to check the unchosen box $B$
from 1 and 2, and indeed, by Observation \ref{obs:alg2mat}, $(2/11) / 0.4 = (3/11) / 0.6 = 5/11$, which is the probability of not checking $B$ given that it was not checked up to this point.

In this section we present Algorithm $\Anew$, which given $p$, calculates the function $L$, and randomly chooses boxes so as to get $L$ as its function. 
We describe the ideas behind it here, and the formal proof appears in appendix \ref{apx:optimal}.
\thmOptimal
\renewcommand{\thealgorithm}{}
\begin{algorithm*}
\caption{\!\!\!\! \boldmath$\Anew$}
\algblockdefx{From}{EndFrom}{{\bf from }}{}
\algnotext[From]{EndFrom}
\begin{algorithmic}
\For{$t \gets 1$ to $M$}
\For{$y \gets \Active(t-1)+1$ to $\infty$} \Comment{Calculate $\Active(t), \alpha(t)$}
\If{$\sum_{x=1}^y 1 - q(x)/q(y) > t$}
\State $\Active(t) \gets y-1$
\State $\alpha(t) \gets (\Active(t) - t)/\sum_{x \leq \Active(t)} q(x)$ 
\EndIf
\EndFor 
\From unchecked boxes $x\leq \Active(t)$ \Comment{Choose one box}
\If{$x \leq \Active(t-1)$}
\State Check $x$ w.p.\ $1 - \alpha(t)/\alpha(t-1)$
\Else
\State Check $x$ w.p.\ $1 - \alpha(t)q(x)$
\EndIf
\EndFrom
\EndFor
\end{algorithmic}
\end{algorithm*}

At step $t$, the first thing $\Anew$ does is calculate the values of $L(x,t)$ for all $x$, so that it can recreate them with its random choices. For that it needs to calculate $\alpha(t)$, which by Observation \ref{obsExact} means solve the equation:
\[[eq:alpha]
t = \sum_x 1 - L(x,t) = \sum_x 1 - \min(1, \alpha(t)q(x))
\]
The first step is to figure out which $x$'s actually contribute something to this sum. 
Say box $x$ is \emph{active} at time $t$ if $L(x,t) < 1$. As $L$ is non-decreasing in $x$, there is some $\Active(t)$, s.t.\ the set of active boxes at time $t$ is $\set{1, \ldots, \Active(t)}$.
To calculate $\Active(t)$, $\Anew$ gradually decreases $\alpha(t)$, while keeping the column requirement satisfied. The point is, $x$ is active when $\alpha(t) < 1/q(x)$, and so to see who is active, it needs to only check $\alpha(t) = 1/q(1), 1/q(2), \ldots$. Once $\Active(t)$ is found, solving \eqref{eq:alpha} and finding $\alpha(t)$ is straightforward.

Now that $L(x,t)$ is calculated, $\Anew$ randomly chooses a box to check according to it, using the fact that up to this point, the probability that box $x$ was not checked is $L(x,t-1)$. 
If a box was not active, and now is, then clearly it should be checked with probability $1-q(x)\alpha(t)$. 
If it was already active, then it should change from $q(x)\alpha(t-1)$ to $q(x)\alpha(t)$, which by Observation \ref{obs:alg2mat} means it should be checked with probability $1 - \alpha(t)/\alpha(t-1)$. Fortunately, all these probabilities sum up to $1$.

As an interesting side note, observe that at each step, all previously active yet unchecked boxes get the same probability of being checked. Moreover, this probability does not depend at all at the previous choices made by the algorithm. This point sounds counter-intuitive from a Bayesian point of view, as we would expect a rescaling of the probabilities that differs according to the history we've already seen.

An important point is that $\Anew$ has at each step a finite set of boxes to choose from. As $p$ goes to $0$, $q$ goes to infinity, and so if there are an infinite number of active boxes, then $\alpha$ must be $0$, but that means that all boxes were surely checked.

How does algorithm $\Anew$ look for example distributions, and how does it compare to $\Aold$?
In general it is quite difficult to analyse the exact running time of this algorithm, but sometimes it can be done, as we shall see.

\subsection{Uniform Distribution}

The first example that comes to mind is when the treasure is uniformly placed in one of the boxes $\set{1, \ldots, M}$. 
As $q(x)$ is equal for all boxes, an agent running $\Anew$ will at the first step choose among them uniformly, and continue to do so at each step, choosing from those that it did not check yet.
This algorithm is the most natural choice in this case, and indeed, by Theorem \ref{thmOptimal} it is optimal. Analysis is simple:
\[\BA
\Time(\Anew) 
& =
\sum_{t=0}^M \prob{\text{not found by time $t$}}
=
\sum_{t=0}^M \prod_{i=0}^{t-1} \B{1 - \F{1}{M-i}}^k
=
\sum_{t=0}^M \prod_{i=0}^{t-1} \BF{M - i - 1}{M- i}^k
\\ & =
\sum_{t=0}^M \BF{M - t}{M}^k
=
\R{M^k} \sum_{i=0}^M i^k 
\approx
\R{M^k} \F{M^{k+1}}{k+1}
=
\F{M}{k+1}
\EA\]
Note that with coordination, the expected running time would be about $M/2k$, so we lose about a factor of $2$ by non-coordination as opposed to $4$ in the case of Algorithm $\Aold$.
This algorithm is memory intensive, yet if we choose to simplify and just choose uniformly at random a box from all boxes at each step, we get that the running time is practically the same for large $M$:
\[\BA
\sum_{t=0}^\infty \B{1 - \R{M}}^{kt}
=
\R{1 - \B{1 - \R{M}}^k}
\approx
\F{M}{k}
\EA\]

%

\section{Pareto Distributions}
$\Anew$ is optimal, but it is a complex algorithm. For a large family of Pareto distributions we present a simplified algorithm that approximates the performance of $\Anew$ well. 
Let $r_{b,M}$ be the Pareto distribution with parameter $b>0$ on $M$ boxes. Denote $b(x) = 1/x^b$, and then $r_{b,M}(x) = I / b(x)$, where $I = 1/\sum_{x=1}^M b(x)$ is the normalization factor. Note that the function $b(\cdot)$ will be important on its own right. We will especially be interested in the case\footnote{In fact, our lower bound result also hold for $b=1$, but our upper bound proof does not work for this case. However, we strongly believe the theorem to be true for $b=1$ as well.
} where $b < 1$, as when $M$ grows, the fraction of the weight any specific box has goes to $0$. For $b>1$ that is not true, and so we are left with too little leeway for simplifying $\Anew$.

In Algorithm $\Apareto$, each agent, at its $t$-th step, chooses uniformly from one of the boxes it did not check yet in $\set{1, \ldots, \min(M, \Floor{t/\sigma})}$, where $\sigma = b/(b + k - 1)$. While $\Apareto$ is not optimal, asymptotically it is. Practically all proofs of the section appear in Appendix \ref{apx:lowerPareto}. 
\thmPareto
In what follows, $o(1)$ means an expression that tends to $0$ as $M$ goes to infinity.
\subsection{Lower Bound}

The lower bound part of Theorem \ref{thmPareto} is proved for all non-coordinating algorithms. For that, instead of the set of functions in $\Valid$, we consider a more general class of functions and so lower bound the original question. For a measurable set $X$ denote:
\[
\Cont(X) = \stset{N : X \times [0,\infty] \rightarrow [0,1]}{N(\cdot, t) \text{ is measurable for every fixed } t}
\]
For an $N \in \Cont(X)$, we say that $N$ satisfies the {\em column requirements} if for all $t$:
$
\column_N(t) = \int_X 1-N(x,t) \dx \leq t 
$.
Such a function is called {\em valid}, and $\Valid(X)$ is the set of all valid functions.
Given an integer $k \geq 2$ and some measurable function $p : X \rightarrow [0, \infty)$, define:
\[\BA
\Utime_{p, k}(N)
= 
\int_0^\infty  \int_X p(x) N(x,t)^k \dx \dt
\EA\]
This is a sort of equivalent of the $\Time$ of algorithms, but is ``unnormalized", as  $p$ is not necessarily a distribution.
The following claim shows a connection between algorithms and functions:
\rClaim{clmMatFunc}{
For every distribution $p$ on $\{1,2,\ldots,M\}$ and algorithm $A$ on the $M$ boxes, there is a function $N \in \Valid([1,M+1])$ such that  $\Utime_{p', k}(N) \leq \Time_{p, k}(A)$, where 
$p': [1,M+1] \rightarrow [0,\infty)$ is any non-increasing measurable function that agrees with $p$.
}
It is proved quite directly by taking $N(x,t) = A(\Floor{x}, \Floor{t})$. This shows that lower bounding the ``running time" of functions in $\Valid([1,M+1])$ will lower bound the running time of algorithms on $M$ boxes.
Next, fix some $0 < b < 1$, and so the function $b(x) = 1/x^b$.
\rObservation{obsOpt}{
Let $X$ be a finite interval of $\Real^+$. 
Among all functions of $N \in \Valid(X)$ there is one that minimizes $\Utime_{b,k}(N)$. 
Denote it $\OPT_{b,X}$.
}
The proof of this observation uses the full power of Lemma \ref{lmMain} by finding the optimal function of $x$ for each specific $t$, in a very similar way to the optimality proof of $\Anew$. 
Next, we introduce the important tool of \emph{zooming}, which is used a couple of times in what follows.
\begin{definition}
Given some $N \in \Cont(X)$ and $u,v > 0$, define the zooming of $N$ by $(u,v)$ as: 
$
\zoom{N}{u}{v}(x,t) = N(x/u, t/v)
$,
where $\zoom{N}{u}{v}(x,t) \in \Cont(uX)$. 
\end{definition}
The intuitive meaning of it is that the algorithm is expanded to work on a domain of size $u$ times the original one, and slowed down by a factor of $v$. 
What happens to the column requirement integrals and to the time?
\rLemma{lmZoom}{
For $N \in \Cont(X)$ and $u,v > 0$,  
$
\Utime(\zoom{N}{u}{v}) = u^{1-b}v \Utime(N)
$, and for all $t$, 
$
\column_{\zoom{N}{u}{v}}(t) = u \column_N\BF{t}{v}
$.
}
This Lemma reduces our question to the running time of a specific set of optimal functions:
\rClaim{clmReduction}{
For any Algorithm $A$ that works on $M$ boxes, denoting $r = r_{b,M}$, and $\epsilon = 1/(M+1)$:
\[
\F{\Time_r(A)}{\Time_r(\Acord)} \geq (1 - o(1)) \cdot  
k(2-b) \cdot \Utime_b(\OPT_{[\epsilon, 1]})
\]
}
To use Claim \ref{clmReduction} one should figure out who is $\OPT_{[\epsilon, 1]}$. This is possible using Lemma \ref{lmMain}, but the equations that calculate $\alpha(t)$ are differential and it is not clear how to solve them. 
However, assuming $M$ is large, we can trick our way out of this via a clever use of zooming, and so reduce the problem to calculating $\OPT_{(0,1]}$ which is much simpler. Denote $\OPT = \OPT_{(0,1]}$. Then:
\rClaim{clmEpsilon}{
$
\lim_{\epsilon \rightarrow 0} \B{
\Utime(\OPT_{[\epsilon, 1]}) / 
\Utime(\OPT)}
= 1
$
}
All that is left to do, is figure out $\OPT$ and calculate its running time:
\rClaim{clmOPTvalue}{
Denote $\sigma = b/(b+k-1)$. Then,  
$\Utime(\OPT) = \F{\sigma(2-\sigma)}{2 -b} + \F{(1-\sigma)^2}{k+1}$.
}
Finally, we can prove the lower bound and optimality part of Theorem \ref{thmPareto}.
By Claim \ref{clmReduction}, Claim \ref{clmEpsilon} and Claim \ref{clmOPTvalue}, for every algorithm $A$:
\[
\lim_{M \rightarrow \infty}
\F{\Time(A)}{\Time(\Acord)}
\geq 
k(2-b) 
\lim_{M \rightarrow \infty}
\Utime(\OPT_{[1/(M+1), 1]}) 
=
k(2-b) 
\Utime(\OPT)
 =
k\sigma(2-\sigma) + \F{k(2-b)(1-\sigma)^2}{k+1}
\]

\subsection{Upper Bound}

Below we describe the high level structure of the proof of the upper bound part of Theorem \ref{thmPareto}. The missing proofs of this section appear in Appendix \ref{apx:upperPareto}. 
A simple analysis of $\Acord$ and gives:
\rClaim{clmUpperStart}{
\[
\F{\Time(\Apareto)}{\Time(\Acord)}
\leq
\F{k(2-b)}{M^{2-b}}
\sum_t \sum_{x=1}^M \R{x^b} \Apareto(x,t)^k
\]
}
Since $\Apareto$ chooses uniformly from a set of unopened boxes at each stage, by Observation \ref{obs:alg2mat}, when $x$ is in this set then:
\[
\Apareto(x,t) = \Apareto(x,t-1) \cdot \B{1 - \R{|\text{interval chosen from}| - (t-1) }}
\]
Applying generously and then using Lemma \ref{lmGamma}, one gets:
\rClaim{clmUpperOne}{
\[
\Apareto(x,t) \leq (1 + o(1)) \cdot 
\begin{cases}
1 & t < \Ceil{\sigma x} \\
\BF{\Ceil{\sigma x}}{t}^\F{b}{k-1} & \Ceil{\sigma x} \leq t <\Ceil{\sigma M} \\
\R{1-\sigma}\B{1 - \F{t}{M}} \BF{\Ceil{\sigma x}}{\sigma M}^\F{b}{k-1}  & \Ceil{\sigma M} \leq t < M \\
0 & t \geq M
\end{cases}
\]
}
The point of this is that completely ignoring the rounding up operations, this is exactly $\OPT(x/M, t/M)$, 
which appears in explicit form in \eqref{eq:opt} of Claim \ref{clmOPTvalue}.
Indeed, using very careful needlework math to get rid of these roundings, we show what would otherwise be a simple claim:
\rClaim{clmUpperHardcore}{
\[
\R{M^{2-b}}
\sum_{t=0}^M \sum_{x=1}^M \R{x^b} \Apareto(x,t)^k
\leq (1 + o(1))
\Utime(\OPT)
\]
}
\noindent Plugging this into Claim \ref{clmUpperStart} gives:
\[
\F{\Time(\Apareto)}{\Time(\Acord)}
\leq
(1 + o(1)) k(2-b) \Utime(\OPT)
\]
Claim \ref{clmOPTvalue} gives the value of $\Utime(\OPT)$, and concludes the upper bound proof of Theorem \ref{thmPareto} in exactly the same fashion as the end of the lower bound proof of this theorem.

\clearpage
\bibliographystyle{plain}
\bibliography{bib}

\clearpage

\section*{\centerline{\Huge Appendix}}
\appendix
\bigskip

\section{Coordinating Agents} \label{apx:coordinating}

\begin{observation} \label{obs:coordinating}
$\Acord$ is optimal among coordinating algorithms, and
$\Time(\Acord) = \sum_x p(x) \Ceil{\F{x}{k}}$.
\end{observation}
\BPF
Coordinating $k$ agents can be viewed as one algorithm that works in phases, where in each phase it can check $k$ boxes. The aim is then to minimize the expected number of phases until the treasure is found. 
In this scenario, as there is no feedback from the algorithm's choices until the treasure
is found, any randomized strategy can be seen as a distribution over deterministic algorithms. It follows then, that it is enough to consider deterministic algorithms. 

W.l.o.g, as this cannot harm the running time, each box is checked once and in each phase there are exactly $k$ boxes that are being checked. Now,
\[
\Time(A) = \sum_x p(x) T(x)
\] 
Where $T(x)$ is the phase where $x$ is checked.
By the assumption above, the sequence $T(1), T(2), \ldots$ contains exactly $k$ copies of each positive integer (representing a phase number). 
Since $p(x)$ is a non-increasing sequence, then by the rearrangement inequality, $\Time(A)$ is minimized when the $T(x)$ are arranged in a non-decreasing order, which is exactly algorithm $\Acord$.
Its running time is clear from definitions.
\EPF

\section{Observation \ref{obsExact}} \label{apx:obsExact}

\obsExact
\BPF
For $t \geq M$, $\alpha(t) = 0$ satisfies the column requirement, and is as required.
Next assume that $0<t<M$. Clearly, in this case $\alpha(t) \neq 0$, as otherwise the column requirement is violated. 
Assume by contradiction that $C_L(t) \neq t$, and
since $L$ is valid this means that $C_L(t) < t$. 

Note that since $p(x)$ goes to zero, $q(x)$ goes to infinity, and so there are only a finite number of $x$'s where $\alpha(t)q(x) < 2$. As $\alpha(t) > 0$, we can reduce it slightly, and this will only affect the value of $L$ at these $x$'s. Making this change small enough, will maintain the inequality $C_L(t) < t$, and keep $L$ valid. As this change can only decrease $L(x,t)$ at these points, 
$\Time(L)$ does not increase. Contradicting the minimality of $\alpha(t)$.
\EPF

\section{Refined Analysis of the Algorithms of 
	\texorpdfstring{\cite{fraigniaud_parallel_2016}}{Fraignaud et al.}}
\label{apx:refined}

\subsection{Efficiency of $\Aold$} \label{apx:refinedUniversal}

\clmRefined
\BPF
Count the time in steps of size $2$, so at each step $\Aold$ chooses two new boxes. The algorithm might actually end mid-step, but this just means that this is an over approximation.

The number of elements the algorithm chooses from at step $t$ is 
$(k+1)t - 2(t-1) = (k-1)t + 2$. 
Box $x$ starts to have some probability of being checked at time $s = \Ceil{x/(k+1)}$, and for 
$t \geq s$ the probability of $x$ not being checked by time $t$ is:
\[
\prod_{i=s}^t \B{1 - \F{2}{(k-1)i + 2}}^k
=
\prod_{i=s}^t \BF{(k-1)i}{(k-1)i+2}^k
 = 
\prod_{i=s}^t \BF{i}{i+\F{2}{k-1}}^k
\leq
\BF{s+1}{t}^\F{2k}{k-1}
\]
Where the last step is by Claim \ref{clm:gamma2} proved in Appendix \ref{apx:gamma} below. 
Denoting $a = 2k/(k-1)$ the total running time for $x$ is  then at most (times $2$):
\[
s + 2 + \sum_{t=s+2}^\infty \BF{s+1}{t}^a
\]
As $((s+1)/t)^a$ is decreasing, we can bound the sum from above by taking the integral but starting it at $s+1$ and not $s+2$. This gives the upper bound of:
\[\BA
&
s + 2 + \int_{s+1}^\infty \BF{s+1}{t}^a \dt
= 
s + 2 + (s+1) \int_1^\infty t^{-a} \dt
=
s + 2 + \F{s+1}{a-1} 
\\ & =
1 + (s+1)\B{1 + \R{\F{2k}{k-1}-1}}
=
1 + \B{\Ceil{\F{x}{k+1}}+1}\B{1 + \F{k-1}{k+1}}
\\ & \leq
1 + \B{\F{x}{k+1}+2}\BF{2k}{k+1}
\leq
5 + \F{2k}{(k+1)^2}s
\EA\]
Multipying by $2$ gives the result.
\EPF

\subsection{Efficiency of $\Amem$} \label{apx:refinedMemory}

\clmRefinedMemory
\BPF
The proof proceeds in very similar to that of Theorem \ref{clmRefined}, yet is in fact a little simpler. Count the time in steps of size $2$.

Box $x$ starts to have some probability of being checked at time $s = \Ceil{x/k}$, and for 
$t \geq s$ the probability of $x$ not being checked by time $t$ is:
\[
\prod_{i=s}^t \B{\B{1 - \F{1}{ki}}^2}^k
\leq
\prod_{i=s}^t \B{1 - \R{ki +1}}^{2k}
= 
\prod_{i=s}^t \BF{i}{i+\F{1}{k}}^{2k}
\leq
\BF{s}{t}^2
\]
Where the last step is by Lemma \ref{lmGamma} below. 
The total running time for $x$ is then at most (times $2$):
\[
s + 1 + \sum_{t=s+1}^\infty \BF{s}{t}^2
\]
As $(s/t)^2$ is decreasing, we can bound the sum from above by taking the integral but starting it at $s$ and not $s+1$. This gives the upper bound of:
\[\BA
&
s + 1 + \int_{s}^\infty \BF{s}{t}^2 \dt
= 
s + 1 + s \int_1^\infty \R{t^2} \dt
=
2s + 1
\EA\]
Multipying by $2$ gives the result.
\EPF

\subsection{Gamma Function Property} \label{apx:gamma}

\begin{claim} \label{clm:gamma2}
For integers $b \geq a \geq 1$, and $k\geq 2$,
\[
\prod_{i=a}^b \F{i}{i+\F{2}{k-1}} \leq \BF{a+1}{b}^\F{2}{k-1}
\]
\end{claim}
\BPF
Let us start with the case $k=2$. In this case, $2/(k-1) = 2$.
If $a = b$ then this is clearly true, as left side is at most $1$, and right side at least $1$.
Otherwise, the product is telescopic and we get:
\[
\prod_{i=a}^b \F{i}{i+2} = \F{a(a+1)}{(b+1)(b+2)}
\]
It is indeed at most $(a+1)^2/b^2$, as its numerator is smaller, and its denominator larger.

Regarding $k \geq 3$. In this case, $2/(k-1) \leq 1$, and so we can use Lemma \ref{lmGamma} below, to get that the product in the claim is at most:
\[
\BF{a}{b}^\F{2}{k-1} \leq \BF{a+1}{b}^\F{2}{k-1} 
\]
\EPF
\lmGamma
\BPF
By induction on $a$ (somehow on $b$ it doesn't work..).
If $b = a$, then we should show 
$a/(a + \phi) \leq 1$, which is true.
We therefore assume that the result holds for $a+1$ and prove it for $a$:
\[
\prod_{i=a}^b \F{i}{i+\phi}
= \F{a}{a+\phi} \cdot \prod_{i=a+1}^b \F{i}{i+\phi} 
\leq \F{a}{a+\phi} \cdot \BF{a+1}{b}^\phi
\] 
We want to show:
\[
 \F{a}{a+\phi} \cdot \BF{a+1}{b}^\phi  \leq \BF{a}{b}^\phi 
\quad \Longleftrightarrow \quad
\F{a}{a+\phi}  \leq \BF{a}{a+1}^\phi   
\quad \Longleftrightarrow \quad
\B{\F{a+\phi}{a}}^\R{\phi}  \geq \B{\F{a+1}{a}}
\]
Take $b=\R{a}<1$ and $x=\R{\phi}\geq 1$ the above is equivalent to:
\[
\B{1+\F{b}{x}}^x  \geq \B{1+b}
\]
If we show that the left side is increasing with $x$ when $x\geq 1$ then we are done.
We take the derivative (using an internet site):
\[
\B{1+\F{b}{x}}^x \cdot \left( \ln\B{1+\F{b}{x}}  - \F{b}{x\B{1+\F{b}{x}}}\right)
\]
This is positive if
\[
\B{1+\F{b}{x}}\ln\B{1+\F{b}{x}}  > \F{b}{x}
\]
Take $y=\F{b}{x}\leq 1$. We want to show that 
\[
\B{1+y}\ln\B{1+y}  > y
\]
We use the equality
\[
\ln\B{1+y} = \int_0^y \R{1+t}\dt
\]
So 
\[
\B{1+y}\ln\B{1+y} = \int_0^y \F{1+y}{1+t}\dt> \int_0^y  1 \dt =y
\]
 as desired.
\EPF

\section{Proof of Main Lemma} \label{apx:mainLemma}

Recall that $V(T)$ the set of measurable functions $f : X \rightarrow [0,1]$ such that $\int 1 - f \dm \leq T$. Also,
for a measurable function $c: X \rightarrow [0,\infty)$, and $\alpha \geq 0$ the function $f_{c, \alpha} : X \rightarrow [0,1]$ is:
\[\BA
f_{c, \alpha}(x) = \begin{cases}
1 & c(x) = 0 \\
\min\B{1, \alpha c(x)^{-\R{k-1}}} & \text{otherwise}
\end{cases}
\EA\]
In what follows we will drop the subscript $c$ when it is clear from context.

\lmMain
\BPF
A few of points to begin with:
\BE
\I
All of the functions below are measurable, either by definition or by straightforward proof. Also, as all of them are positive, they will all have a defined Lebesgue integral (though its value may be $\infty$.
\I
Starting from the end, assuming that the existence of $\alpha$ is proved, we claim that
$\alpha = \min\stset{\beta \geq 0}{f_\beta \in V(T)}$. Existence proves that this set is not empty. By monotonicity, if $\beta < \beta'$ then 
$\int c f^k_\beta \dm \leq \int c f^k_{\beta'} \dm$, so all that remains to show is that this minimum exists. If not, then there is some sequence $\set{\beta_n}_{n=1}^\infty$ that approaches an infimum $\alpha$. By the definitions, $f_{\beta_n}$ converges pointwise to $f_\alpha$. By Fatou's lemma:
\[
\int 1 - f_\alpha \dm \leq \liminf_{n \rightarrow \infty} \int 1 - f_{\beta_n} \dm \leq T
\]  
So $f_\alpha \in V(T)$, proving this point. 
\I \label{eq:TisSmall}
Denote $S = \stset{x \in X}{c(x) > 0}$, the support of $c$.
If $T \geq \mu(S)$, then take $\alpha = 0$. We get $f_0(x) = 0$ on $S$ and $1$ elsewhere,
so $\int 1 - f_0 \dm = \mu(S) \leq T$, and so $f_0 \in V(T)$. Also, $\int c f_0^k \dm = 0$ and  is therefore optimal, so we are done.
We will therefore always assume that $T < \mu(S)$. Specifically, $\mu(S) > 0$ and $T < \mu(X)$.
\I \label{eq:conditionA}
For any $\epsilon > 0$, examine the set $Y = \stset{x \in X}{c(x) > \epsilon}$.
We claim that $\mu(Y) < \infty$. 
Denote $Z = \stset{x \in X}{h(x) < 1/2}$. As $h \in V(T)$, 
\[
T \geq \int 1 - h \dm \geq \int_Z \R{2} \dm \geq \F{\mu(Z)}{2}  
\]
So $\mu(Z) < \infty$, and therefore $\mu(Y \cap Z) < \infty$. Also:
\[
\infty > \int c h^k \dm 
\geq
\int_{Y \cap \neg Z} c h^k \dm
\geq
\F{\epsilon}{2^k} \cdot \mu(Y \cap \neg Z)
\]
Together, this means that $\mu(Y) < \infty$.
\EE
The proof proceeds as usual in these cases, by a gradual increase of the generality of the function $c$ that we handle.

\paragraph{Indicator functions.}
First assume $c = 1_A$ is the indicator function of some set $A \subseteq X$.
By Item \ref{eq:conditionA}, $\mu(A) < \infty$,
and by Item \ref{eq:TisSmall}, we can assume $T < \mu(A)$.
For any $g \in V(T)$:
\[\BA
\int c g^k \dm & = 
\int_A g^k \dm 
\geq \mu(A) \cdot \B{ \R{\mu(A)}\int_A g \dm }^k 
\\ & 
\geq \R{\mu(A)^{k-1}} \cdot \B{\mu(A) - \int_A{1 - g}\dm}^k 
\geq \F{\B{ \mu(A) - T }^k}{\mu(A)^{k-1}}
\EA\]
Where we used Jensen's inequality for the case where the total measure is not necessarily $1$.
Take $\alpha = (\mu(A) - T)/\mu(A)$. We get $0 < \alpha < 1$, and
$f_\alpha(x) = \alpha$ for every $x\in A$ and $1$ elsewhere. Also, 
\[
\int 1 - f_\alpha \dm = \int_A 1 - f_\alpha \dm = \int_A \F{T}{\mu(A)} \dm = T
\]
So $f_\alpha \in V(T)$.
Also,
\[
\int c f_{\alpha}^k \dm = \int_A \alpha^k \dm = \F{(\mu(A) - T)^k}{\mu(A)^{k-1}}
\]
And so $f_{\alpha}$ is optimal. Note that it is a constant function on $A$.

\paragraph{Simple functions.} 
In this case, $c = \sum_{i=1}^n c_i 1_{X_i}$, where all $c_i > 0$, and the $X_i$ are pair-wise disjoint. Also, by Item \ref{eq:conditionA}, all the $X_i$ are of finite measure. 

Given some $g \in V(T)$, let us examine it on each of the $X_i$'s separately. Denote $T_i = \int_{X_i} 1 - g \dm$. Restricted to $X_i$, according to the case of indicator functions, there is some constant $g_i \geq 0$, such that 
$\int_{X_i} 1 - g_i \dm \leq T_i$, and
$\int_{X_i} g_i^k \dm \leq \int_{X_i} g^k \dm$.

We therefore define $g' = 1_Y + \sum_{i=1}^n g_i 1_{X_i}$, where $Y = X \setminus \cup_{i=1}^n X_i$.
According to the above, 
\[
\int 1 - g' \dm = \sum_{i=1}^n \int_{X_i} 1 - g_i \dm \leq \sum_{i=1}^n \int_{X_i} 1 - g \dm
\leq
\int 1 - g \dm \leq T
\] 
So $g' \in V(T)$. Also, 
\[
\int c g^k \dm
=
\sum_{i=1}^n c_i \int_{X_i} g^k \dm
\geq
\sum_{i=1}^n c_i \int_{X_i} g_i^k \dm
=
\int c g'^k \dm 
\]
So $g'$ is a better candidate than $g$, and we can therefore assume that $g$ is constant on each of the $X_i$'s, and can be written as
$g = \sum_{i=1}^n g_i 1_{X_i}$. 

Our question can now be viewed as follows. Given $c_1, \ldots, c_n$ and $\mu_1, \ldots \mu_n > 0$, find the $g_1, \ldots, g_n \in [0,1]$ among those satisfying $\sum_{i=1}^n \mu_i(1 - g_i) \leq T$, that minimize
$\sum_{i=1}^n c_i g_i^k$. As the solution space is compact and the function to minimize is continues, there exists an optimal solution $g_1, \ldots, g_n$.

Take some $i$ such that $1 < i \leq n$. We can rebalance the values of $g_1$ and $g_i$ as we wish, as long as the sum $\mu_1g_1 + \mu_ig_i$ remains the same. According to Lemma \ref{lm:balance2} below, these two values must satisfy:
\[
g_i = \min\B{1, \BF{c_1}{c_i}^\R{k-1} g_1}
\]
Taking $\alpha = c_1^{1/(k-1)} g_1$, we obtain the form $g_i = \min(1, \alpha c_i^{-1/(k-1)})$ which concludes this case.

\paragraph{The general case.}
Let $\set{c_n}_{n=1}^\infty$ be a non-decreasing family of simple functions that have $c$ as their pointwise limit. According to the simple function case, for each $n$ there is some $\alpha_n$ such that the function $f_n = f_{c_n, \alpha_n}$ gives minimal $\int c_n f_n^k \dm$ among all functions of $V(T)$.

If this sequence of $\alpha_n$ is unbounded, we can keep only a sub-sequence where $\alpha_n \rightarrow \infty$ and define 
$f(x) = \lim_{n\rightarrow \infty} f_n(x) = 1$ everywhere.  
Otherwise we can keep only a converging sub-sequence of the $\alpha_n$, and denote its limit by $\alpha$. Now, define the function
$f(x) = f_{c, \alpha}(x) = \lim_{n\rightarrow \infty} f_n(x)$. 
Either way the pointwise limit of the $f_n$'s exists and we denote it by $f$.

Examine the sequence of functions $1 - f_n$. They all satisfy 
$\int 1 - f_n \dm \leq T$, and so by Fatou's lemma:
\[
\int 1 - f \dm
 \leq \liminf_{n\rightarrow \infty} \int 1 - f_n \dm \leq T
\]
And so $f \in V(T)$.
For all $x$, the function $c f^k$ is the pointwise limit of $c_n f_n^k$. 
As $f_n$ is optimal for $c_n$,
\[
\int c_n f_n^k \dm 
\leq
\int c_n h^k \dm
\leq
\int c h^k \dm
<
\infty
\]
So all these integrals are jointly bounded and so their $\liminf$ exists. Therefore, by Fatou's lemma:
\[
\int c f^k \dm \leq \liminf_{n \rightarrow \infty} \int c_n f_n^k \dm < \infty
\]
Assume there is some $g$ that is better than $f$. That is, there is some $\delta > 0$ such that:
\[
\int c g^k \dm < \int c f^k \dm - \delta
\]
Take large enough $n$, and use the fact that $f_n$ is optimal for $c_n$,
\[
\int c f^k \dm - \F{\delta}{2}
< \int c_n f_n^k \dm 
\leq \int c_n g^k \dm 
\leq \int c g^k \dm 
\]
and we get a contradiction.

The only thing left to show is that $f = f_{c,\alpha}$ for some $\alpha$. As we've seen there are two cases, and we have to deal with the case where $f = 1$.

By \eqref{eq:TisSmall} there is some
$\epsilon > 0$ such that the set $A = \stset{x \in X}{c(x) > \epsilon}$ satisfies $\mu(A) > 0$, and by \eqref{eq:conditionA}, $\mu(A) < \infty$.
If $T < \mu(A)$, then
take the function $g(x) = 1 - T/\mu(A)$ on this set and $1$ elsewhere. 
Clearly $g \in V(T)$. Also, 
\[
\int c1^k \dm - \int cg^k \dm = \int_A c \cdot \F{T}{\mu(A)} \dm \geq \epsilon T > 0
\]
As $f$ is optimal, it cannot be the function $1$ and must be of the required form.
If $T > \mu(A)$, proceed in the same way, except $g(x) = 0$ on $A$ and $1$ elsewhere.
\EPF

\subsection{Lemma for 2}

\begin{lemma} \label{lm:balance2}	
Let $k\geq 2$, $c_1, c_2, \mu_1, \mu_2 > 0$, and $M \leq \mu_1 + \mu_2$.	
The minimal value of $\mu_1 c_1 g_1^k + \mu_2 c_2 g_2^k$, where $g_1, g_2 \in [0,1]$ and $\mu_1 g_1 + \mu_2 g_2 = M$ is	
achieved only when: 
\[
g_1 = \min\B{1, \B{c_2/c_1}^\R{k-1} \cdot g_2}
\]
\end{lemma}
\BPF
Let $c = c_2/c_1$, and $\mu = \mu_2 / \mu_1$. Setting $N = M/\mu_1$, we can write the lemma equivalently as follows. Assuming $N \leq 1 + \mu$, knowing that $g_1 + \mu g_2 = N$, the $g_1,g_2 \in [0,1]$ minimizing $g_1^k + c \mu g_2^k$ satisfy $g_1 = \min\B{1, c^{1/(k-1)} g_2}$.

Denoting $g_2 = (N - g_1)/\mu$, we want to minimize:
\[
g_1^k + \F{c}{\mu^{k-1}}(N - g_1)^k
\]
We take the derivative w.r.t.\ $g_1$:
\[[eq:firstDerivative]
k\B{ 
g_1^{k-1} - \F{c}{\mu^{k-1}} (N - g_1)^{k-1}
}
\]
This is zero exactly when:
\[[eq:minx]
g_1 = c^\R{k-1} \cdot \F{N - g_1}{\mu} \B{ = c^\R{k-1} g_2 }
\]
We get:
\[[eq:positive]
g_1 =  \F{c^\R{k-1}}{\mu + c^\R{k-1}} N
\]
We take the second derivative (the first was \eqref{eq:firstDerivative}),
\[
k(k-1)\B{g_1^{k-2} + \F{c}{\mu^{k-1}} (N-g_1)^{k-2}}
\]
If we look at $g_1$'s in the range $[0,N]$, this is always strictly positive, meaning our function is U shaped there. Also, by \eqref{eq:positive} the minimum is somewhere in $[0,N]$. Recall that $g_1 \in [0,1]$. If the bottom of the U is in~$[0,1]$ then as we've seen in \eqref{eq:minx} we get the lemma. 
Otherwise it must be somewhere in $(1,N]$, and so our minimum would be at $g_1 = 1$. Note that it is unique.
\EPF

\section{Optimality proof of \texorpdfstring{$\Anew$}{A star}} \label{apx:optimal}

\thmOptimal
\BPF
First, $\Anew$ calculates $\alpha(t)$ and $\Active(t)$.
Note that $y \leq \Active(t)$ iff $\alpha(t) < 1/q(y)$, and so, to calculate $\Active(t)$, it is enough to check values for $\alpha(t)$ that are equal to $1/q(y)$ for $y > \Active(t-1)$. 
Once we know $\Active(t)$, by Observation \ref{obsExact}:
\[
t = \sum_{x \leq \Active(t)} 1 - \alpha(t) q(x)
\]
Solving this for $\alpha(t)$ is what the algorithm does.

To show that the next part of $\Anew$ is at all valid, we show that the probabilities of each step add up to at most $1$.
The number of boxes that were already active at $t-1$, and were not checked yet at time $t$ is
$\Active(t-1) - (t-1)$. So, summing all the probabilities of the different boxes:
\[[eq:sumProbs]
(\Active(t-1) - t + 1)\B{1 - \F{\alpha(t)}{\alpha(t-1)}} + 
\sum_{\Active(t-1) < x \leq \Active(t)} 1 - \alpha(t)q(x) 
\]
By Observation \ref{obsExact}:
\[
\sum_{x \leq \Active(t-1)} 1 - \alpha(t-1)q(x) = t - 1
\RIGHT
 \sum_{x \leq \Active(t-1)} \alpha(t-1) q(x) = \Active(t-1) - t + 1
\]
Plugging this is \eqref{eq:sumProbs}:
\[\BA
&
\sum_{x \leq \Active(t-1)} (\alpha(t-1) - \alpha(t))q(x)
+
\sum_{\Active(t-1) < x \leq \Active(t)} 1 - \alpha(t)q(x) 
\\ & = 
\sum_{x \leq \Active(t)} 1 - \alpha(t)q(x)
-
\sum_{x \leq \Active(t-1)} 1 - \alpha(t-1)q(x)
\EA\]
By Observation \ref{obsExact} the first sum is $t$ and the second is $t-1$, and so the sum of probabilities is indeed $1$.

The last bit is to show that indeed $\Anew = L$. This is proved by induction on $t$. For $t = 0$, $L(x, 1) = \Anew(x, 1)$ for all $x$. Assume equality for $t-1$ and we prove it for $t$. 
For $x \leq \Active(t-1)$, $\Anew(x, t-1) = L(x,t-1) = \alpha(t-1)q(x)$. Using Observation \ref{obs:alg2mat}: 
\[
\Anew(x, t) = 
\Anew(x, t-1) \cdot \F{\alpha(t)}{\alpha(t-1)}
 = 
\alpha(t-1)q(x) \cdot \F{\alpha(t)}{\alpha(t-1)} = L(x, t)
\]
For $\Active(t-1) < x \leq \Active(t)$, it is straightforward.
\EPF

\section{Lower Bounding Pareto Distributions} \label{apx:lowerPareto}

\subsection{Claim \ref{clmMatFunc}}
\clmMatFunc
\BPF
Define $N(x,t) = A(\Floor{x},\Floor{t})$.
For any $t$:
\[
\column_N(t) 
= 
\int_1^{M+1} 1 - A(\Floor{x},\Floor{t}) \dx
=
\sum_{x=1}^M 1 - A(x, \lfloor t \rfloor)
= \column_{A}(\Floor{t}) \leq 
\Floor{t} \leq t
\]
So $N$ satisfies the column requirements. Next,
\[\BA
\Utime_{p', k}(N) 
& = 
\int_0^\infty \int_1^{M+1}  p'(x) A(\Floor{x} ,\Floor{t})^k \dx \dt
\leq
\int_0^\infty \int_1^{M+1} p(\Floor{x}) A(\Floor{x}, \Floor{t})^k \dx \dt
\\ & =
\sum_{x=1}^M \sum_{t=0}^\infty p(x) A(x, t) = \Time_{p, k}(A)
\EA\]
\EPF

\subsection{Observation \ref{obsOpt}}
\obsOpt
\BPF
Fix $t$. Setting $c(x) = b(x)$ and $T=t$, Lemma \ref{lmMain} gives a function $f_t(x)$ minimizing 
$\int_X b(x) f_t(x)^k \dx$, under the condition $\int_X 1 - f_t(x) \dx \leq t$. 
To fulfil the condition of the lemma, take $h(x) = 0$ for $x\leq t$, and $1$ elsewhere. Then $\int_X 1 - h(x) \dx \leq t$, and as $X$ is a finite interval. Also, 
\[
\int_X \R{x^b}h(x)^k \dx = \int_{X \setminus [0,t]} \R{x^b} \dx < \infty
\]
Putting all these $t$'s together by setting $\OPT_{b,X}(x,t) = f_t(x)$, we get that $\OPT_{b,X} \in \Valid(X)$. Also:
\[
\Utime(\OPT_{b,X}) = \int_0^\infty \int_X \R{x^b} f_t(x)^k \dx
\] 
And as the $f_t$ minimize the inner integral for any function in $\Valid(X)$, we get that if this integral exists it is minimal. 
To show it exists, we note that the function $F(t) = \int_X \R{x^b} f_t(x)^k \dx$ is non-increasing in $t$, and so is measurable. This is because, if $t < t'$, $f_t \in V(t')$, and so by minimality of $f_{t'}$, 
$F(f_{t'}) \leq F(f_t)$. This means that $\Utime(\OPT_{b,X})$ is defined, although it might be $\infty$.
\EPF

\subsection{Zooming Lemma}
\lmZoom
\BPF
First:
\[
\Utime(\zoom{N}{u}{v}) 
= 
\int_0^\infty \int_{uX} \R{x^b} N(x/u,t/v)^k \dx \dt
=
uv \int_0^\infty \int_X  \R{(xu)^b} N(x, t)^k \dx \dt
= 
u^{1-b}v \Utime(N)
\]
The column integrals:
\[
C_{\zoom{N}{u}{v}}(t)  = 
\int_{uX} 1-N(x/u, t/v)\dx 
= 
u \int_X 1-N(x, t/v)\dx = u C_N\BF{t}{v}
\]
\EPF

\subsection{Claim \ref{clmReduction}}

\clmReduction
\BPF
Recall $r(x) = I/x^b$, where $I = 1/\sum_{i=1}^M 1/x^b$. We take $r'$ to be the extension of this on all of $[1,M+1]$. Then,  
by Claim \ref{clmMatFunc} there is some $N \in \Valid([1, M+1])$ such that: 
\[
\Time_r(A) \geq \Utime_{r'}(N) = I \cdot \Utime_b(N)
\]
Where the last step is trivial when examining the definition of $\Utime$.
Consider $N' = \zoom{N}{\epsilon}{\epsilon}$. By Lemma \ref{lmZoom} and the fact that $N$ satisfies the column requirements,  
\[
C_{N'}(t) = \epsilon C_N\BF{t}{\epsilon} \leq \epsilon \cdot \F{t}{\epsilon} = t 
\]
So $N' \in \Valid([\epsilon, 1])$. Also, by the same lemma,
\[
\Utime_b(N') = \R{(M+1)^{2-b}} \Utime_b(N)
\]
Together with the fact that $\Utime_b(N') \geq \Utime_b(\OPT_{[\epsilon,1]})$, we obtain:
\[
\Time_r(A) \geq (M+1)^{2-b} \cdot I \cdot \Utime_b(\OPT_{[\epsilon,1]})
\]
The running time of $\Acord$ is:
\[\BA
\Time_r(\Acord) 
& =
\sum_{x=1}^M \F{I}{x^b} \Ceil{\F{x}{k}} 
\leq
\sum_{x=1}^M \F{I}{x^b} \B{\F{x}{k} + 1}
=
\F{I}{k} \sum_{x=1}^M x^{1-b} + 1
\\ & \leq
1 + \F{I}{k} \int_1^{M+1} x^{1-b} \dx 
=
1 + \F{I}{k} \F{(M+1)^{2-b}}{2-b} 
\EA\]
Where we used the fact that $x^{1-b}$ is monotonically non-decreasing. Together:
\[
\F{\Time_r(A)}{\Time_r(\Acord)} 
\geq 
\F{(M+1)^{2-b} \cdot I \cdot \Utime_b(\OPT_{[\epsilon,1]})}{1 + \F{I}{k} \F{(M+1)^{2-b}}{2-b}}
=
\F{1}{\F{k (2-b)}{I (M+1)^{2-b}} + 1} \cdot 
k (2-b) \Utime_b(\OPT_{[\epsilon,1]})
\]
As $b > 0$, then $I^{-1} = \sum_{i=1}^M 1/x^b = o(M)$, and so the first factor tends to $1$ as $M$ tends to infinity.
\EPF

\subsection{Getting Rid of \texorpdfstring{$\epsilon$}{Epsilon}}

\clmEpsilon
\BPF
For the sake of this proof, 
denote $E = \OPT_{[\epsilon, 1]}$.
To show that the limit is at most $1$, define $\OPT'$ to be $\OPT$ restricted to $[\epsilon, 1]$.
It is easy to see that $\OPT' \in \Valid([\epsilon, 1])$, and so $\Utime(\OPT') \geq \Utime(E)$. 
Also, it is clear that $\Utime(\OPT') \leq \Utime(\OPT)$, which concludes this side.

To show that the limit is indeed $1$,
we construct a new function $E'$ that will span the whole range of $x$'s from $0$ to $1$, with little change to $\Utime(E)$.
This will be done by slowing $E$ down, and using what we saved in the column integrals to visit the $x$'s between $0$ and $\epsilon$ using our optimal solution, running it fast enough so it does not incur a big difference in $\Utime(E)$.

Fix some $a < 1$ to be determined later. Define:
\[
E'(x,t) = 
\begin{cases}
\zoom{\OPT}{\epsilon}{\epsilon/(1-a)} & x \leq \epsilon \\
\zoom{E}{1}{1/a} & x > \epsilon
\end{cases}
\]
Since the zoomed version of $\OPT$ here is defined on the $x$'s in $(0,\epsilon]$ and the zoomed $E$ is on those in $[\epsilon, 1]$, we get that for all $t$:  
\[
\column_{E'}(t) 
= 
\column_{\zoom{\OPT}{\epsilon}{\epsilon/(1-a)}}(t) + \column_{\zoom{E}{1}{1/a}}(t) 
= 
\epsilon \column_{\OPT}\B{\F{1-a}{\epsilon} t} + \column_E(at)
  \leq
\epsilon \F{1-a}{\epsilon} t + at = t
\]
where we used Lemma \ref{lmZoom} and the fact that both $\OPT$ and $E$ satisfy the column requirements.
So $E' \in \Valid((0,1])$ and by the optimality of $\OPT$, 
$\Utime(E') \geq \Utime(\OPT)$.
Again, by Lemma \ref{lmZoom}:
\[
\Utime(\OPT)
\leq
\Utime(E')
= 
\Utime(\zoom{\OPT}{\epsilon}{\epsilon/(1-a)})
+
\Utime(\zoom{E}{1}{1/a})
=
\F{\epsilon^{2-b}}{1-a} \Utime(\OPT) 
+
\R{a} \Utime(E)
\]
And therefore, denoting $\epsilon' = \epsilon^{2-b}$,
\[\BA
\Utime(E) \geq a\B{1 - \F{\epsilon'}{1-a}} \Utime(\OPT)
\EA\]
Taking $a = 1 - \sqrt{\epsilon'}$, the factor becomes:
$\B{1 - \sqrt{\epsilon'}}^2$, 
which goes to $1$ as $\epsilon$ goes to $0$.
\EPF

\subsection{Analysing $\OPT$}

\clmOPTvalue
\BPF
We proceed as in Observation \ref{obsOpt}, and use Lemma \ref{lmMain} to figure out the exact structure of $\OPT$.
For each fixed $t$ it gives the function $f_t(x)$ minimizing
$\int_0^1 \R{x^b} f_t(x)^k \dx$, from all those satisfying the column requirement $\int_0^1 1 - f_t(x) \leq t$. 
As shown in the observation's proof, $\OPT(x,t) = f_t(x)$. 

The first step is to understand what is $f_t(x)$. 
For $t \geq 1$, the optimal $f_t$ is obviously $f_t(x) = 0$, since this satisfies the column requirement and has an integral of $0$.
For $t<1$,  we have $\int_0^1 1 - f_t(x)\dx \leq t$,
and that $f_t(x) = \min(1, \alpha_t x^{b/(k-1)})$. So, given $t$, it is possible to deduce what its corresponding $\alpha$ is (we drop the subscript).
For each $t$, denote by $\gamma$ (a function of $t$) the smallest $x$ where $f_t(x) = 1$, and in case this does not happen, set $\gamma = 1$. So:
\[[eq:gamma]
\gamma = \min\B{1, \R{\alpha^\F{k-1}b}}
\]
for every $t<1$, to minimize our target function, we would like $f_t$ to be the smallest possible and so the column requirement will actually be an equality:
\[
t = \int_0^1 1 - \min\B{1, \alpha x^\F{b}{k-1}} \dx
= \gamma - \int_0^\gamma \alpha x^\F{b}{k-1} \dx
\]
We have two cases:
\BE
\I
for all $t$ where $\gamma < 1$ this equation is:
\[[eq:gamma-t]
\gamma - t = \int_0^\gamma \alpha x^\F{b}{k-1} \dx
\]
From \eqref{eq:gamma}, and using the assumption that $\gamma<1$, we get $\alpha = 1/\gamma^{b/(k-1)}$. Plugging this in:
\[
\gamma -t 
=
\int_0^\gamma \left(\F{x}{\gamma}\right)^\F{b}{k-1} \dx 
= 
\gamma \int_0^1 x^\F{b}{k-1} \dx 
= 
\gamma \R{\F{b}{k-1} + 1} 
= 
\F{k-1}{b + k-1} \gamma
\]
Recall $\sigma = b/(b+k-1)$, and so $\gamma = t/\sigma$.
This means, that for all $t < \sigma$, $\gamma < 1$ and then 
$\alpha = (\sigma/t)^{b/(k-1)}$. For all other $t$, $\gamma = 1$.
\I
for all $t$ where $\gamma = 1$:
\[
1 - t 
= \int_0^1 \alpha x^\F{b}{k-1} \dx
= \alpha \R{\F{b}{k-1} + 1} = (1 - \sigma) \alpha 
\]
and so:
\[
\alpha = \F{1-t}{1-\sigma}
\]
\EE
Putting all this together:
\[[eq:opt]
\OPT(x,t) = \begin{cases}
1 & t \leq \sigma x \\
\BF{\sigma x}{t}^\F{b}{k-1} & \sigma x < t \leq \sigma \\
\F{1-t}{1-\sigma} x^\F{b}{k-1} & \sigma < t \leq 1 \\
0 & t > 1
\end{cases}
\]

Next, we wish to calculate $\Utime(\OPT)$.
\[\BA
\Utime(\OPT)
& =
\int_0^1 \int_0^1 \R{x^b} \OPT(x,t)^k \dx \dt 
\\ & = 
\int_0^{\sigma} \left( \int_0^\gamma \R{x^b}(\alpha x^\F{b}{k-1})^k \dx \right) \dt
+    
\int_0^{\sigma} \left( \int_\gamma^1 \R{x^b}1^k \dx \right) \dt 
+
\int_{\sigma}^1 \left(  \int_0^1 \R{x^b}(\alpha x^\F{b}{k-1})^k \dx  \right) \dt
\EA\]
Focus on each term separately:
\BE
\I
Here $\gamma = t/\sigma < 1$. Hence, the inner integral is:
\[
\int_0^\gamma \R{x^b}(\alpha x^\F{b}{k-1})^k \dx   
= \alpha^{k-1} \int_0^\gamma \alpha x^\F{b}{k-1} \dx  
 = \alpha^{k-1} ( \gamma - t) = \R{\gamma^b} (\gamma - t)
\]
Where the second equality is due to (\ref{eq:gamma-t}). Then, as $\gamma = t/\sigma$, we get:
\[
\BF{\sigma}{t}^b \B{\F{t}{\sigma} - t} 
= 
\sigma^b t^{1-b} \F{1 - \sigma}{\sigma}
\]
The whole integral if $b<1$:
\[
(1 - \sigma)\int_0^{\sigma}  \BF{t}{\sigma}^{1-b} \dt 
=
\sigma(1 - \sigma)\int_0^1 t^{1-b} \dt 
=
\F{\sigma(1 - \sigma)}{2-b}
\]
If $b=1$ then it is $\sigma(1-\sigma)$ which is the same.
\I
Here, still, $\gamma < 1$. Two cases:
\BE
\I
If $b = 1$, the inner integral is:
\[
\int_\gamma^1 \R{x} \dx = \log(1) - \log(\gamma) = -\log(\gamma) 
\]
Plugging in $\gamma = t/\sigma$, and calculating the whole integral:
\[
\int_0^{\sigma} - \log(t/\sigma) \dt = - \sigma \int_0^1 \log(t)\dt  = \sigma
\]
Last bit is because indefinite integral of $\log(x)$ is $x\log(x) - x$.
\I
If $b<1$:
\[
\int_\gamma^1 \R{x^b} \dx 
= \R{1-b}\B{1 - \gamma^{1-b}}
\]
Plugging in $\gamma = t/\sigma$, and calculating the whole integral:
\[
\R{1-b} \int_0^{\sigma} 1 - \BF{t}{\sigma}^{1-b} \dt
=
\F{\sigma}{1-b} \B{1 - \int_0^{1} t^{1-b} \dt}
= 
\F{\sigma}{1-b} \B{1 - \R{2-b}}
= 
\F{\sigma}{2-b}
\]
\EE
So $\sigma/(2-b)$ works for both cases.
\I
Here $\gamma = 1$.
\[
\int_0^1 \R{x^b}(\alpha x^\F{b}{k-1})^k \dx
= 
\alpha^{k-1} \int_0^1 \alpha x^\F{b}{k-1} \dx  
= 
\alpha^{k-1} (1 - t)
\]
Plugging in $\alpha = (1-t)/(1-\sigma)$ and calculating the whole integral:
\[\BA
&
\R{(1-\sigma)^{k-1}}\int_{\sigma}^1 (1-t)^k  \dt 
=
\R{(1-\sigma)^{k-1}}\int_0^{1-\sigma} t^k  \dt 
\\ & =
\R{(k+1)(1-\sigma)^{k-1}} (1-\sigma)^{k+1}  \dt 
 =
\F{(1-\sigma)^2}{k+1}
\EA\] 
\EE 
In total:
\[\BA
\F{\sigma(1-\sigma)}{2 -b} + \F{\sigma}{2-b} + \F{(1-\sigma)^2}{k+1}
=
\F{\sigma(2-\sigma)}{2 -b} + \F{(1-\sigma)^2}{k+1}
\EA\]
\EPF

\section{Upper Bounding Pareto Distributions} \label{apx:upperPareto}

\subsection{Bounding the ratio $\Time(\Apareto) / \Time(\Acord)$}

\clmUpperStart
\BPF
Setting $I = (\sum_{x=1}^M 1/x^b)^{-1}$, 
the running time of $\Acord$ is (noting that $x^{1-b}$ is non-decreasing):
\[
\Time(\Acord) 
=
\sum_{x=1}^M \F{I}{x^b} \Ceil{\F{x}{k}} 
\geq
\F{I}{k} \sum_{x=1}^M x^{1-b}
\geq
\F{I}{k} \int_0^M x^{1-b} \dx 
=
\F{I}{k} \F{M^{2-b}}{2-b} 
\]
So:
\[
\F{\Time(\Apareto)}{\Time(\Acord)}
=
\F{\sum_t \sum_{x=1}^M \F{I}{x^b} \Apareto(x,t)^k}
{\Time(\Acord)}
\leq
\F{k(2-b)}{M^{2-b}}
\sum_t \sum_{x=1}^M \R{x^b} \Apareto(x,t)^k
\]
\EPF

\subsection{Figuring out $\Apareto$'s Function}

\clmUpperOne
\begin{proof}
As mentioned, since $\Apareto$ chooses uniformly from a set of unopened boxes at each stage, by Observation \ref{obs:alg2mat}, when $x$ is in this set then:
\[
\Apareto(x,t) = \Apareto(x,t-1) \cdot \B{1 - \R{|\text{interval chosen from}| - (t-1) }}
\]
Also,
\BE
\I
Fix $x$. When $x > \Floor{t/\sigma}$ it has no probability of being checked, and as $x$ is an integer, this means $x > t/\sigma$, and so $t < \sigma x$. It therefore starts being checked when $t = \Ceil{\sigma x}$.
\I
The checking is over all unchosen boxes when $M \leq \Floor{t/\sigma} \leq t/\sigma$. This starts when $t = \Ceil{M\sigma}$.
\EE
Combining all this together gives:
\[
\Apareto(x,t) \leq
\begin{cases}
1 & t < \Ceil{\sigma x} \\
\prod_{i=\Ceil{\sigma x}}^t \B{1 - \R{i/\sigma - i + 1}} & \Ceil{\sigma x} \leq t < \Ceil{\sigma M} \\
\prod_{i=\Ceil{\sigma x}}^{\Ceil{\sigma M}-1} \B{1 - \R{ i/\sigma - i + 1}} 
\prod_{i=\Ceil{\sigma M}}^t \B{1 - \R{M - i+1}} & \Ceil{\sigma M} \leq t < M \\
0  & t \geq M
\end{cases}
\]
Where $i/\sigma$ replaces $\Floor{i/\sigma}$ in the probabilities, as it only increases the result.
Now,
\[\BA
\prod_{i=\Ceil{\sigma x}}^t \B{1 - \R{i/\sigma - i + 1}}
=
\prod_{i=\Ceil{\sigma x}}^t \F{(1/\sigma - 1)i }{(1/\sigma - 1)i + 1}
= 
\prod_{i=\Ceil{\sigma x}}^t \F{i}{i + \F{\sigma}{1-\sigma}}
\leq
\BF{\Ceil{\sigma x}}{t}^\F{b}{k-1}
\EA\]
Where Lemma \ref{lmGamma} is used for last step. Similarly:
\[\BA
& \prod_{i=\Ceil{\sigma M}}^t \B{1 - \R{M - i+1}}
 = 
\prod_{i=\Ceil{\sigma M}}^t \F{M-i}{M-i+1}
=
\F{M - \Ceil{\sigma M}}{M - \Ceil{\sigma M}+1} \cdots \F{M - t }{M - t + 1}
\\ & = 
\F{M-t}{M - \Ceil{\sigma M}+1}
\leq
\F{M-t}{M - \sigma M} 
=
\R{1-\sigma}\B{1 - \F{t}{M}}
\EA\]
So:
\[
\Apareto(x,t) \leq \begin{cases}
1 & t < \Ceil{\sigma x} \\
\BF{\Ceil{\sigma x}}{t}^\F{b}{k-1} & \Ceil{\sigma x} \leq t <\Ceil{\sigma M} \\
\R{1-\sigma}\B{1 - \F{t}{M}} \BF{\Ceil{\sigma x}}{\Ceil{\sigma M}-1}^\F{b}{k-1}  & \Ceil{\sigma M} \leq t < M \\
0 & t \geq M
\end{cases}
\]
Finally, multiply the third case by $\B{(\Ceil{\sigma M} - 1) / \sigma M}^{b/(k-1)}$. This will decrease the final result by at most this factor, which tends to $1$ as $M$ goes to infinity, giving the result.
\end{proof}

\subsection{Relating $\Apareto$ and $\OPT$}

\clmUpperHardcore
\BPF
Our aim is to show:
\[[eq:temp]
\sum_{t=0}^M \sum_{x=1}^M \R{x^b} \Apareto(x,t)^k
\leq 
M^{2-b} \int_0^1 \int_0^1 \R{x^b} \OPT(x,t)^k \dx \dt
\]
while being quite loose in this comparison, as we can allow additive terms of $o(M^{2-b})$ and still get the result.
Since $\OPT$ is non-increasing in $t$, then so is $\int_0^1 \R{x^b} \OPT(x,t)^k \dx$. The right side is then at least:
\[
M^{1-b} \sum^M_{t=1} \int_0^1 \R{x^b} \OPT(x,t/M)^k \dx
\]
The case $t=0$ contributes at most an additive $\sum_{x=1}^M 1/x^b = o(M)$ to the left side of \ref{eq:temp}, and so is insignificant. This is in fact true for any particular $t$. So to prove \eqref{eq:temp}, we will show that for all but a small constant number of $t$'s:
\[[eq:upperTarget1]
\sum_{x=1}^M \R{x^b} \Apareto(x,t)^k
\leq 
M^{1-b} \int_0^1 \R{x^b} \OPT(x,t/M)^k \dx
\]
Where here, additive terms of order $o(M^{1-b})$ are considered insignificant. 

Putting side by side $\Apareto$'s upper bound as presented in Claim \ref{clmUpperOne} (ignoring the $1 + o(1)$ factor which does not bother us), and the explicit form of $\OPT$ of \eqref{eq:opt}, shows their resemblance:
\[
\begin{array}{lcr}
\begin{cases}
1 & t < \Ceil{\sigma x} \\
\BF{\Ceil{\sigma x}}{t}^\F{b}{k-1} & \Ceil{\sigma x} \leq t <\Ceil{\sigma M} \\
\R{1-\sigma}\B{1 - \F{t}{M}} \BF{\Ceil{\sigma x}}{\sigma M}^\F{b}{k-1}  & \Ceil{\sigma M} \leq t < M \\
0 & t \geq M
\end{cases}
& \quad \quad \quad \quad \quad \quad &
\begin{cases}
1 & t \leq \sigma x \\
\BF{\sigma x}{t}^\F{b}{k-1} & \sigma x < t \leq \sigma \\
\F{1-t}{1-\sigma} x^\F{b}{k-1} & \sigma < t \leq 1 \\
0 & t > 1
\end{cases}
\end{array}
\]
Fix some $t$, and set $f(x) = \OPT(x/\sigma, t/M)^k$. 
Clearly $\OPT(x, t/M)^k = f(\sigma x)$, but also:
\[
\Apareto(x, t)^k \leq \OPT\B{\F{\Ceil{\sigma x}}{\sigma M}, \F t M}^k =   f\BF{\Ceil{\sigma x}}{M}
\]
for all but possibly $t = \Ceil{\sigma M} \pm 1$.
So to prove \eqref{eq:upperTarget1}, it will be enough to prove that for any function $f$ where $f(\cdot) \in [0,1]$:
\[[eq:upperTarget]
\sum_{x=1}^M \R{x^b}f\BF{\Ceil{\sigma x}}{M} \leq M^{1-b}\int_0^1 \R{x^b}f(\sigma x) \dx
\]
Assuming the integral above is defined.
Again, additive terms of order $o(M^{1-b})$ are considered insignificant.
The next step is to approximate the integral by a very specific Riemann sum. This gives a result that is correct up to a multiplicative factor that tends to $1$, which is fine for our purpose. 
The $n$-th interval is $I_n = (i_{n-1}/M, i_n/M]$, where
$i_n = \Ceil{n/\sigma}$.
$I_n$ is sampled at $n/\sigma M $, which is clearly an inner point of $I_n$. 

For example, if $\sigma = 0.3$ then  $I_1 = (1, 4], I_2 = (4, 7], I_3 = (7,10],
I_4 = (10, 14]$, and the respective sample points are $3\R3, 6\F23, 10$ and $13\R3$. Of course, all of this should be divided by $M$.
Note that the size of the intervals is about $1/M\sigma$ and so tends to $0$, as required.

The right hand side of \eqref{eq:upperTarget} is approximated by:
\[
M^{1-b} \cdot \R{M} \sum_{n=1}^{\Ceil{\sigma M}} (i_n - i_{n-1}) \R{(n/\sigma M)^b} f(n/M)
=
\sum_{n=1}^{\Ceil{\sigma M}} \B{i_n - i_{n-1}} \F{f(n/M)}{(n/\sigma)^b}
\]
This can be written as:
\[[eq:midSum]
\sum_{x=1}^M
\F{f(n_x/M)}{(n_x/\sigma)^b}
\]
Where $n_x$ is defined to satisfy  $x \in (i_{n_x-1}, i_{n_x}]$, so that indeed, each
term in the original sum appears exactly $i_n - i_{n-1}$ times in the new sum (of course without the factor of $(i_n - i_{n-1})$).
The condition on $n_x$ is actually
$\Ceil{\F {n_x-1} \sigma} < x \leq \Ceil{\F{n_x} \sigma}$.
The following is proved in Appendix \ref{apx:rounding}:
\rObservation{obsRound}{
If $x$ and $n$ are natural numbers, then 
\[
\Ceil{\F {n-1} \sigma} < x \leq \Ceil{\F n \sigma}
\LEFTRIGHT
n = \Ceil{\sigma x}
\]
}
Using this observation, \eqref{eq:midSum} is:
\[
\sum_{x=1}^M \R{ (\Ceil{\sigma x} / \sigma)^b} f\BF{\Ceil{\sigma x}}{M}
\geq
\sum_{x=1}^M \R{ (x + \R{\sigma})^b} f\BF{\Ceil{\sigma x}}{M}
\]
Since $b\leq 1$, $x^b$ is sub-linear, and so:
\[
\R{x^b} - \R{ (x + \R{\sigma})^b}
=
\F{(x + \R{\sigma})^b - x^b}{x^b(x + \R\sigma)^b}
\leq
\F{\R{\sigma}}{x^b(x + \R\sigma)^b}
\leq
\R{\sigma x^{1+b}}
\]
Noting that $f(\cdot) \leq 1$, the above is at least:
\[
\sum_{x=1}^M \R{x^b} f\BF{\Ceil{\sigma x}}{M}
-
\R{\sigma} \sum_{x=1}^M \R{x^{1+b}}
\]
If $b<1$, then the second term is bounded above by some constant independent of $M$, and so in this case is $o(M^{1-b})$, proving \eqref{eq:upperTarget}.
\EPF

\subsubsection{Proof of Rounding Observation} \label{apx:rounding}

\obsRound 
\BPF
The case where $n = \sigma x$ is straightforward. Otherwise, 
\BE
\I
\[ 
n = \Ceil{\sigma x}
\RIGHT n > \sigma x
\RIGHT x < \F n \sigma 
\RIGHT x < \Ceil{\F n \sigma}
\]
Also,
\[ 
n = \Ceil{\sigma x}
\RIGHT n -1  < \sigma x
\RIGHT \F{n-1}\sigma < x
\RIGHT \Ceil{\F{n-1}\sigma} \leq x
\]
But equality would mean that $n-1 = \sigma x$, which cannot be true.
\I
\[
x < \Ceil{ \F n \sigma }
\RIGHT x < \F n \sigma
\RIGHT \sigma x < n
\]
And,
\[
x > \Ceil{ \F{n-1}\sigma }
\RIGHT x > \F{n-1}\sigma
\RIGHT \sigma x > n-1
\]
So $\Ceil{\sigma x} = n$.
\EE
\EPF

\end{document}